\documentclass{osa-article}
\usepackage{color}
\usepackage{amsmath}
\usepackage{bm}

\renewcommand{\Re}{\operatorname{Re}}
\renewcommand{\Im}{\operatorname{Im}}

\journal{oe}
\articletype{Research Article}

\begin{document}

\title{Teleportation-based noiseless quantum amplification of coherent states of light}

\author{Jarom{\' i}r Fiur{\' a}{\v s}ek\authormark{*}}

\address{Department of Optics, Palack\'{y} University, 17. listopadu 1192/12, 771 46 Olomouc, Czech Republic\\

\email{\authormark{*}fiurasek@optics.upol.cz}
}

\begin{abstract}
We propose and theoretically analyze a teleportation-based scheme for high-fidelity noiseless quantum amplification of coherent states of light. In our approach, the probabilistic noiseless quantum 
amplification operation is encoded into a suitable auxiliary two-mode entangled state 
and then applied to the input coherent state via continuous-variable quantum teleportation.
The scheme requires conditioning on the outcomes of homodyne measurements in the teleportation protocol. In contrast to high-fidelity noiseless quantum amplifiers 
based on combination of conditional single-photon addition and subtraction, 
the present scheme requires only photon subtraction in combination with auxiliary Gaussian  squeezed vacuum states. 
We first provide a pure-state description of the protocol which allows us to to clearly explain its principles and functioning. Next we develop a more comprehensive model based on phase-space representation of quantum states, 
that accounts for various experimental imperfections such as excess noise in the auxiliary squeezed states or limited efficiency of the single-photon detectors that can only distinguish the presence or absence of photons. 
We present and analyze predictions of this phase-space model of the noiseless teleamplifier.
\end{abstract}

\section{Introduction}

The laws of quantum physics impose fundamental limits on the performance of optical amplifiers. In particular, phase insensitive amplification of light unavoidably adds noise to the amplified signal. 
This can be understood by observing that the hypothetical noiseless amplification of coherent states $|\alpha\rangle\rightarrow |g\alpha\rangle$
with gain $g>1$ decreases the overlap between the output states and therefore it cannot be a deterministic physical  operation, i.e. a trace-preserving completely positive map.
These limits of deterministic amplifiers can be overcome by probabilistic amplifiers \cite{Pandey2013}, where successful amplification is heralded by 
certain outcome of measurement on an auxiliary quantum system. The probabilistic noiseless amplifiers approximate the (still unphysical) conditional operation $g^{\hat{n}}$, where $\hat{n}$ 
is the photon number operator, that maps input coherent state $|\alpha\rangle$ onto an amplified coherent state, 
$\hat{g}^{n}|\alpha\rangle=e^{(g^2-1)|\alpha|^2/2}|g\alpha\rangle$. The heralded noiseless quantum amplifiers were theoretically proposed by Ralph and Lund \cite{Ralph2009} and soon 
afterwards experimentally demonstrated by several groups \cite{Xiang2010,Ferreyrol2010,Usuga2010,Zavatta2011,Osorio2012,Kocsis2013}. 
The  probabilistic noiseless quantum amplifiers have attracted significant attention \cite{Fiurasek2009,Marek2010,Fiurasek2012,Walk2013,Chrzanowski2014,Haw2016,McMahon2014} because they can be used for 
 suppression of  losses in quantum optical communication \cite{Gisin2010,Ralph2011,Micuda2012}, entanglement distillation \cite{Bernu2014,Adnane2019}, improvement of continuous-variable quantum teleportation \cite{Adnane2019tel}, 
or breeding of larger highly non-classical Schr\"{o}dinger cat-like states formed by superpositions of coherent states.

The original noiseless quantum amplifier for weak coherent states $|\alpha\rangle$ was based on the modified quantum scissors scheme \cite{Pegg1998}, 
which truncates the Fock space to the subspace of vacuum and single-photon states. This, however, limits the performance of the amplifier and an amplification gain larger than $1$ can be achieved 
only for coherent states with amplitude $|\alpha|<0.5$. For larger coherent states one can either split the signal into several modes and utilize several elementary noiseless amplifiers in parallel \cite{Ralph2009}, or 
consider a generalized quantum scissors scheme with higher Fock-state ancilla $|n\rangle$ \cite{Winnel2020,Guanzon2021}, which however rapidly increases the experimental complexity of the amplifier. 
To avoid these drawbacks, an alternative approach to noiseless amplification of light was developed  based on combination of single-photon addition and subtraction \cite{Fiurasek2009,Marek2010,Zavatta2011}. 
The resulting conditional noiseless amplifier modulates  the amplitudes of Fock states and the full Fock space is preserved. With this approach, high-fidelity noiseless amplification of coherent states is achievable. 

The single-photon subtraction can be relatively easily  
 realized by tapping off  a small portion of the incoming signal with a highly unbalanced beam splitter with small reflectance $R\ll 1$ and conditioning on detection of a photon in the deflected beam \cite{Ourjoumtsev2006,Nielsen2006,Wakui2007,Kumar2013}, see Fig. 1(a). 
The single-photon addition can be experimentally implemented by injecting the light into the input signal port of a nonlinear crystal that serves as a two-mode squeezer 
operating in the weak squeezing limit \cite{Zavatta2004,Kumar2013}, see Fig. 1(b). Conditioning on detection of a photon 
in the output idler port, we can conclude that a single photon has been coherently added to the signal beam. This scheme has the advantage that inefficient 
single-photon detection only reduces the success rate but does not reduce the quality of photon addition operation, provided that the nonlinear interaction in the crystal is sufficiently weak. 

The photon addition is much more experimentally demanding than photon subtraction and one can ask if a noiseless quantum amplifier can be designed where photon addition is avoided and only photon subtraction is employed. 
It was proposed and experimentally demonstrated that one can approximate noiseless amplification by random displacements of the input coherent state followed by subtraction of several photons from the displaced state. 
This works because the photon subtraction happens with highest probability if the random displacement occurs in the direction of the input coherent state \cite{Marek2010,Usuga2010}. 
However, this process does add some noise to the output state. Here we are interested in implementation of a truly noiseless amplifier, that under perfect circumstances conditionally 
transforms pure input coherent states $|\alpha\rangle$ onto pure output states closely approximating $|g\alpha\rangle$.

\begin{figure}[!t!]
\centerline{\includegraphics[width=0.8\linewidth]{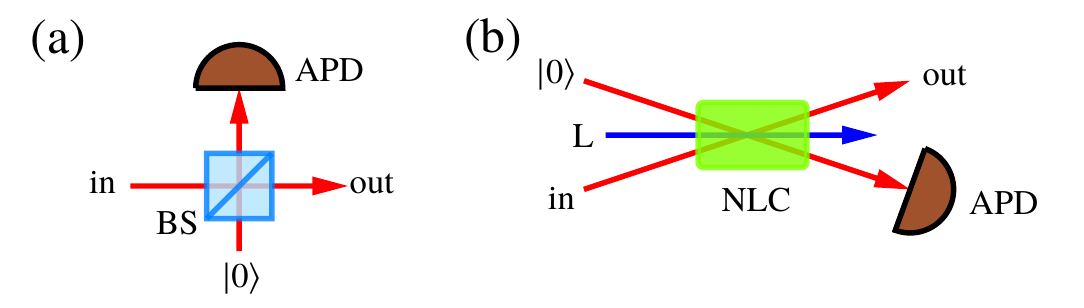}}
\caption{
(a) Conditional single-photon subtraction. BS - highly unbalanced beam splitter with small reflectance $R \ll 1$. APD - single photon detector. (b) Conditional single-photon addition. 
NLC- nonlinear crystal where the input signal and idler modes are coupled by weak two-mode squeezing interaction. L - pump laser beam.  Successful photon addition or subtraction is heralded by click of the detector APD.}
\label{fig01}
\end{figure}

In the present paper, we exploit the concept of teleportation of quantum gates \cite{Gottesman1999} to design a noiseless quantum amplifier based on auxiliary Gaussian squeezed states and photon subtraction. 
The conditional noiseless amplification operation is encoded into suitably modified 
entangled two-mode squeezed vacuum state that serves as a quantum channel in continuous-variable quantum teleportation \cite{Braunstein1998}. 
In contrast to ordinary deterministic quantum teleportation, we must condition on the outcomes of homodyne measurements in the teleportation protocol \cite{Fuwa2014}  to achieve noiseless amplification. 
In this setting we can avoid  the photon addition operation because,  due to the perfect photon number correlations between the signal and idler modes of two-mode squeezed vacuum, 
addition of a photon to the signal mode becomes equivalent to photon subtraction from the idler mode. We thus replace the photon addition operation 
with an ancilla two-mode entangled state that can be prepared from Gaussian two-mode squeezed vacuum states by conditional photon subtractions. 

Previous works have investigated entanglement distillation by noiseless amplification of the two-mode squeezed vacuum state and the improvement of continuous-variable teleportation 
by utilizing the noiselessly amplified two-mode squeezed vacuum state as a quantum channel in teleportation 
\cite{Bernu2014,Adnane2019,Adnane2019tel}.
 Here, we make important step further beyond these previous studies and use the continuous-variable quantum teleportation scheme for conditional  noiseless amplification of the teleported state. 
We note that also the noiseless quantum amplifier based on the quantum scissors scheme can be interpreted as a quantum tele-amplifier, that requires ancilla Fock states and is driven by projection 
 on a suitable multimode $n$-photon state \cite{Guanzon2021}.
 
The rest of the present paper is organized as follows. In Section 2 we present a pure-state description of the protocol which allows us to easily explain its main working principles. 
Subsequently, in Section 3 we develop a more realistic model of the teleportation-based noiseless amplifier 
that takes into account the various experimental imperfections.  This model is based on the powerful and  well established phase-space techniques, where the states are represented 
by their Wigner functions or Husimi $Q$-functions. Finally, Section 4 contains a brief summary and conclusions.

\section{Teleportation-based noiseless quantum amplifier}

We start with a brief overview of the probabilistic noiseless amplification of coherent states via combination of conditional single-photon addition and subtraction. 
The simplest instance of such noiseless quantum amplifier consists of single-photon addition followed by single-photon subtraction \cite{Fiurasek2009,Marek2010,Zavatta2011}, described by the operator 
\begin{equation}
\hat{a}\hat{a}^\dagger=\hat{n}+1,
\label{aadagger}
\end{equation}
where $\hat{a}$ and $\hat{a}^\dagger$ denote the annihilation and creation operators and $\hat{n}$ stands for the photon number operator. For the sake of presentation simplicity, 
we neglect here for the moment the effects of the reflectance $R$ of the tapping beam splitter in the photon subtraction and the finite squeezing in the nonlinear crystal utilized in the photon addition. 
For weak coherent states with $|\alpha| \ll 1$ we can write $|\alpha\rangle \approx |0\rangle +\alpha |1\rangle$ and after the conditional amplification we  get 
\begin{equation}
(\hat{n}+1)(|0\rangle+\alpha|1\rangle)=|0\rangle+2\alpha|1\rangle,
\end{equation}
hence the state is amplified with amplitude gain $g=2$. More generally, we can consider a coherent superposition of single-photon addition followed by single-photon subtraction 
and single-photon subtraction followed by single-photon addition \cite{Fiurasek2009},
\begin{equation}
\hat{G}=\hat{a}\hat{a}^\dagger+(g-2)\hat{a}^\dagger\hat{a}.
\label{Gdefinition}
\end{equation}
This amplifier achieves gain $g$ for weak coherent states, $\hat{G}(|0\rangle+\alpha|1\rangle)=|0\rangle+g\alpha|1\rangle$.
The conditional operation (\ref{Gdefinition}) can be realized with an interferometric scheme, where photon subtraction is attempted both before and after the photon addition 
and the two subtracted beams are overlapped on a beam splitter before detecting the subtracted photon \cite{Costanzo2017,Parigi2007}. 
This erases the which way information about the origin of the subtracted photon and produces the coherent superposition (\ref{Gdefinition}), where the relative weight of the terms 
$\hat{a}\hat{a}^\dagger$ and $\hat{a}^\dagger\hat{a}$ can be controlled by the splitting ratio of the  beam splitter where the two subtracted beams interfere.

\begin{figure}[!t!]
\centerline{\includegraphics[width=0.6\linewidth]{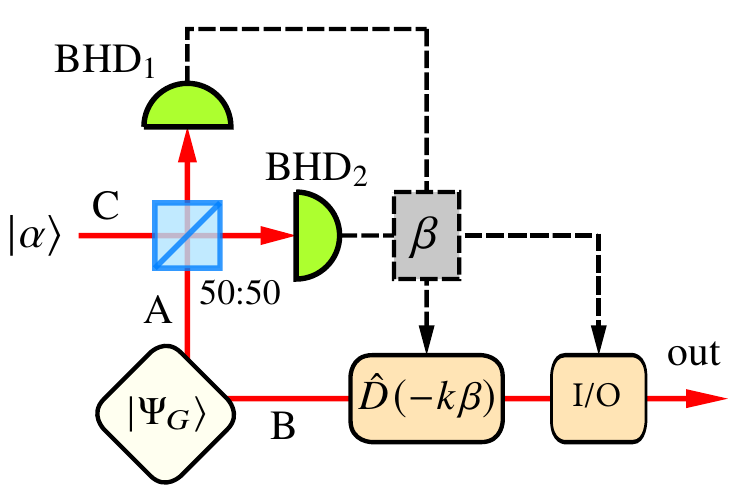}}
\caption{Teleportation-based noiseless quantum amplifier. Modes $A$ and $B$ are prepared in a suitable entangled state $|\Psi_G\rangle$ that encodes the amplification operation $\hat{G}$. 
The input coherent state is injected in mode $C$. Modes $A$ and $C$ interfere on balanced beam splitter and are measured with two balanced 
homodyne detectors BHD.  Mode $B$ can be coherently displaced and the output state is accepted or rejected depending on the measurement outcome $\beta$.}
\label{fig2}
\end{figure}

The conditional photon addition is experimentally much more challenging than the photon subtraction. Moreover, in experiments with continuous signals, the photon subtraction is well mastered, 
while the conditional photon addition to a specific temporal mode may be difficult to accomplish. It is therefore worth asking whether the photon addition 
could be replaced by photon subtraction and other accessible resources. In this paper we answer this question in affirmative by proposing a teleportation-based noiseless quantum amplifier. 
Our approach requires auxiliary Gaussian squeezed states, but it does not require auxiliary single-photon states and we 
avoid injection and mode-matching of the amplified signal beam in a nonlinear crystal. 
Specifically, we can  imprint the conditional operation $\hat{G}$ into a two-mode squeezed vacuum state
\begin{equation}
|\Psi(\lambda)\rangle=\sqrt{1-\lambda^2}\sum_{n=0}^\infty \lambda^n |n,n\rangle,
\label{Psilambda}
\end{equation}
and use the modified state 
\begin{equation}
|\Psi_G(\lambda)\rangle= \frac{1}{\sqrt{P_G}}\hat{I}\otimes\hat{G}|\Psi(\lambda)\rangle
\label{PsiG}
\end{equation}
 as a quantum channel in continuous-variable quantum teleportation.  Here 
\begin{equation}
P_G=\langle \Psi(\lambda)|\hat{I}\otimes \hat{G}^\dagger \hat{G}|\Psi(\lambda)\rangle
\end{equation}
is a normalization factor and $\hat{I}$ denotes a single-mode identity operator.  We utilize the standard Braunstein-Kimble teleportation protocol \cite{Braunstein1998}, see Fig.~2. 
The teleported state is combined on a balanced beam splitter with one part of the two-mode entangled state (\ref{PsiG}) and two homodyne detectors measure the amplitude quadrature $\hat{x}_1$  
of the first output mode and the phase quadrature $\hat{p}_2$ of the second output mode, yielding a complex output signal $\beta=x_1+ip_2$. 
In a deterministic protocol, Bob coherently displaces his mode depending on the measurement outcome $\beta$ to conclude the teleportation. 
However, when we use the modified state $|\Psi_G(\lambda)\rangle$ for teleportation-based noiseless amplification we need to condition on outcomes $\beta$ 
close to $\beta=0$ because the operator $\hat{G}$ does not commute with the displacement operator. Since the  state $|\Psi_G(\lambda)\rangle$ can be prepared before 
any attempted noiseless amplification, the probabilistic nature of the noiseless amplification in this teleportation protocol is solely due to conditioning on the selected range of measurement outcomes $\beta$.

We observe that for $\beta=0$ the noiseless amplification is implemented perfectly up to an additional attenuation due to finite squeezing, i.e. the output state is $\hat{G}|\lambda\alpha \rangle$ 
for any input state $|\alpha\rangle$. To see this, define a  non-normalized infinitely squeezed state
\begin{equation}
|\Psi_{\mathrm{EPR}}\rangle=\frac{1}{\sqrt{\pi}} \sum_{n=0}^\infty |n,n\rangle.
\label{PsiEPR}
\end{equation}
For $\beta=0$, the modes $A$ and $C$  in Fig. 2 are projected on state  $|\Psi_{\mathrm{EPR}}\rangle$, and  the teleported state can be expressed as
\begin{equation}
|\varphi\rangle_B=_{AC}\!\!\langle \Psi_{\mathrm{EPR}}|\Psi_G(\lambda)\rangle_{AB}|\alpha\rangle_C.
\end{equation}
Taking into account the perfect photon number correlations in states $|\Psi(\lambda)\rangle$ and $|\Psi_{\mathrm{EPR}}\rangle$ and expanding the coherent state $|\alpha\rangle$ in Fock basis,
\begin{equation}
|\alpha\rangle=e^{-|\alpha|^2/2}\sum_{n=0}^\infty \frac{\alpha^n}{\sqrt{n!}}|n\rangle,
\end{equation}
we find after some algebra that
\begin{equation}
|\varphi\rangle=\sqrt{\frac{1-\lambda^2}{\pi P_G}} e^{-(1-\lambda^2)|\alpha|^2/2}\hat{G}|\lambda\alpha\rangle.
\end{equation}
We can see that the effective amplification gain is reduced by the finite squeezing strength $\lambda<1$, 
$g_{\mathrm{eff}}=\lambda g$. The performance of the conditional noiseless amplifier can be more comprehensively characterized by the actual amplitude dependent amplification gain
\begin{equation}
g(\alpha)=\frac{1}{\alpha}\frac{\langle \varphi|\hat{a}|\varphi\rangle}{\langle \varphi|\varphi\rangle}=\lambda+\frac{\lambda(g-1)[1+(g-1)|\lambda\alpha|^2]}{[1+(g-1)|\lambda\alpha|^2]^2+(g-1)^2|\lambda\alpha|^2}
\end{equation}
and the fidelity of the output state $|\varphi\rangle$ with the amplified coherent state $|g\lambda \alpha\rangle$,
\begin{equation}
\mathcal{F}(\alpha)=\frac{|\langle \varphi|g\lambda \alpha\rangle|^2}{\langle \varphi|\varphi\rangle}=\frac{[1+g(g-1)|\lambda\alpha|^2]^2}{[1+(g-1)|\lambda\alpha|^2]^2+(g-1)^2|\lambda\alpha|^2}e^{-(g-1)^2|\lambda \alpha|^2}.
\end{equation}

Let us now discuss the preparation of the resource entangled state (\ref{PsiG}). Consider first the noiseless amplification via the conditional operation (\ref{aadagger}). The target entanged state reads
\begin{equation}
\hat{b}\hat{b}^\dagger |\Psi(\lambda)\rangle_{AB} =\sqrt{1-\lambda^2}\sum_{n=0}^\infty (n+1)\lambda^n |n,n\rangle_{AB}.
\end{equation}
Remarkably, this state can be obtained by subtracting a single photon from each mode of the two-mode squeezed vacuum state \cite{Opatrny2000},
\begin{equation}
\hat{a}\hat{b} |\Psi(\lambda)\rangle_{AB}=\lambda \sqrt{1-\lambda^2} \sum_{n=0}^\infty (n+1)\lambda^n|n,n\rangle_{AB}.
\end{equation}
Due to the perfect photon number correlations between the two modes the addition of a photon to mode $B$ becomes equivalent to subtraction of a photon from mode $A$. 
The joint subtraction of two photons from a two-mode squeezed vacuum state has already been successfully demonstrated experimentally \cite{Takahashi2010,Kurochkin2014}.
For weakly squeezed states, the local photon subtractions  can conditionally increase the entanglement of the state and can thus serve for continuous-variable entanglement distillation.

\begin{figure}[!t!]
\centerline{\includegraphics[width=0.4\linewidth]{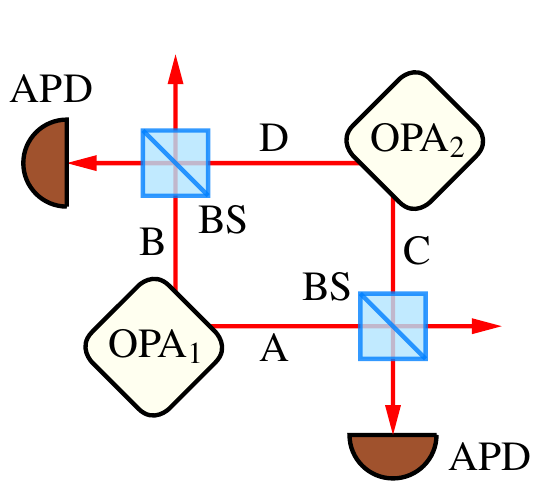}}
\caption{Preparation of entangled state for teleportation-based noiseless amplification via generalized photon subtraction \cite{DellAnno2013}. 
The optical parametric amplifiers OPA generate two-mode squeezed vacuum states with different squeezing strengths $\lambda$ and $\mu$.  
The squeezed states interfere  on beam splitters BS 
and  the output modes $C$ and $D$ are measured with single photon detectors APD. Successful state preparation is heralded by simultaneous click of both detectors APD.}
\label{fig3}
\end{figure}

Consider now preparation of the resource state $|\Psi_G(\lambda)\rangle$ for the more general noiseless amplification operation $\hat{G}$ given by Eq. (\ref{Gdefinition}). We have
\begin{equation}
\left[\hat{b}\hat{b}^\dagger+(g-2) \hat{b}^\dagger\hat{b}\right]|\Psi(\lambda)\rangle_{AB}= \sqrt{1-\lambda^2} \sum_{n=0}^\infty [(g-1)n+1]\lambda^n|n,n\rangle.
\label{Psigeneralizedsubtraction}
\end{equation}
This state can be prepared from a two-mode squeezed vacuum state by a modified joint photon subtraction from each mode, where an auxiliary weak two-mode squeezed vacuum state 
is injected into the auxiliary input ports of the beam splitters that serve for photon subtraction \cite{DellAnno2013}. 
The scheme is illustrated in Fig.~3 and the successful state preparation is heralded by simultaneous click of both single-photon detectors APD. Let us assume that perfect photon number resolving detectors are employed 
that project the output modes $C$ and $D$ on single-photon Fock states $|1,1\rangle_{CD}$. Let $\mu$ denote the squeezing strength of the auxiliary two-mode squeezed vacuum state $|\Psi(\mu)\rangle_{CD}$ and let $T=1-R$ denote the intensity transmittance of the two tapping beam splitters BS in Fig.~3. 
The conditionally prepared state of modes $A$ and $B$ can be expressed as
\begin{equation}
|\Psi_G\rangle_{AB}=\frac{\sqrt{1-\lambda^2}\sqrt{1-\mu^2}}{\sqrt{P_S}}\sum_{n=0}^\infty(T\lambda+R\mu)^{n-1}\left[nRT(\lambda-\mu)^2+(R\lambda+T\mu)(T\lambda+R\mu)\right]|n,n\rangle,
\label{PsiGpreparation}
\end{equation}
where
\begin{equation}
P_S=\frac{(1-\lambda^2)(1-\mu^2)}{(1-\lambda_{\mathrm{eff}}^2)^3}
\left\{
\left[(1-\lambda_{\mathrm{eff}}^2)(R\lambda+T\mu)-\lambda_{\mathrm{eff}}RT(\lambda-\mu)^2\right]^2+R^2T^2(\lambda-\mu)^4
\right\}
\end{equation}
is the probability of successful joint conditional single-photon subtraction. The state (\ref{PsiGpreparation}) is exactly of the form (\ref{Psigeneralizedsubtraction}), with the effective two-mode squeezing constant
\begin{equation}
\lambda_{\mathrm{eff}}=T\lambda+R\mu, 
\end{equation}
nominal amplification gain 
\begin{equation}
g=1+\frac{RT(\lambda-\mu)^2}{(R\lambda+T\mu)(R\mu+T\lambda)},
\label{gformula}
\end{equation}
and a resulting effective amplification gain $g_{\mathrm{eff}}=\lambda_{\mathrm{eff}}g$.

\begin{figure}[!t!]
\centerline{\includegraphics[width=\linewidth]{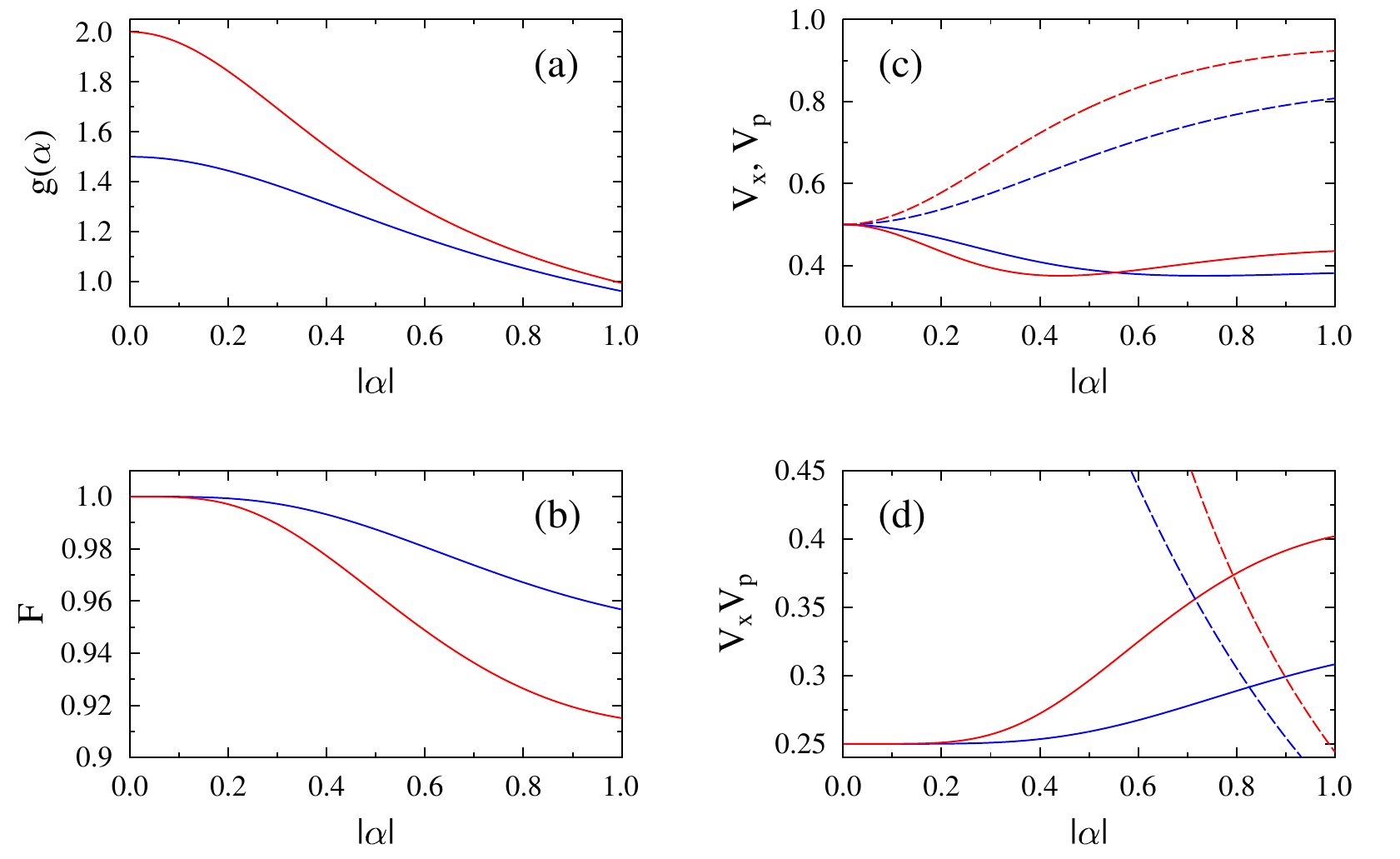}}
\caption{Performance of teleportation-based noiseless quantum amplification of coherent states $|\alpha\rangle$ with pure resource entangled state (\ref{Psigeneralizedsubtraction}) with  $\lambda=0.5$. 
The results are plotted for two different effective gains $g_{\mathrm{eff}}=2$ (red lines) and $g_{\mathrm{eff}}=1.5$ (blue lines). Projection on $\beta=0$ is assumed in the teleportation protocol. 
(a) The amplitude-dependent amplification gain $g(\alpha)$. (b)  Fidelity of the amplified state with coherent state $|g(\alpha)\alpha\rangle$.  (c) Variances $V_x$ (solid lines) and $V_p$ (dashed lines)
of the amplitude and phase quadratures of the amplified state. (d) The uncertainty product $V_xV_p$. The dashed lines show the uncertainty product for the optimal linear deterministic amplifier: $g^2(\alpha)/2-1/4$. }
\label{fig4}
\end{figure}

The experimentally available single-photon detectors usually only distinguish the presence or absence of photons but do not resolve their number. Moreover, the detectors have only limited detection efficiency $\eta$. 
The photon subtraction scheme shown in Fig.~3 can faithfully generate the desired target state (\ref{Psigeneralizedsubtraction}) even 
under such imperfect experimental conditions provided that the beam splitters BS are highly unbalanced and the auxiliary input state $|\Psi(\mu)\rangle_{CD}$ is only very weakly squeezed, 
\begin{equation}
|\mu|^2\ll R \ll 1. 
\label{muinequality}
\end{equation}
Under this condition, the auxiliary input state of modes $C$ and $D$ can be approximated as $|0,0\rangle+\mu|1,1\rangle$. The two dominant events leading to the simultaneous clicks of the two detectors 
APD in Fig.~3 are that either the auxiliary modes $C$ and $D$ are in vacuum state and a single photon is subtracted from both input modes $A$ and $B$ (probability scaling $R^2$), 
or that a single photon is present in each of the auxiliary input modes $C$ and $D$  and these photons are transmitted through the beam splitters BS and detected by APDs (probability scaling $|\mu|^2$).
The dominant unwanted event where one photon is subtracted from mode A and the other APD is triggered by a photon present in the input mode $D$ has a probability that scales as $R|\mu|^2$, which
 is much smaller than the probability of the desired events provided that the inequality (\ref{muinequality}) holds.
 A detailed analysis of the whole teleportation-based noiseless amplifier that takes into account these experimental limitations is provided in the next section. 
Here we just note that in the limit $|\mu|\ll 1$ and $R \ll 1$ the expression for the effective gain simplifies and we can write
\begin{equation}
g_{\mathrm{eff}} \approx T \lambda  \left(1+\frac{R\lambda }{R\lambda +T \mu }\right).
\label{geffformula}
\end{equation}
Note that $\mu$ and $\lambda$ can have opposite signs and therefore the effective gain  $g_{\mathrm{eff}}$ can be arbitrarily high. We observe that the effective amplification gain 
is rather sensitive to small changes of $\mu$. To see this explicitly, we evaluate $dg_{\mathrm{eff}}/d\mu$ and obtain
\begin{equation}
\frac{d g_{\mathrm{eff}}}{d\mu}=2R-\frac{R\lambda^2}{(R\lambda+T\mu)^2}=2R-\frac{\lambda^2(g_{\mathrm{eff}}-\lambda_{\mathrm{eff}})^2}{RT^2(\lambda-\mu)^4}.
\end{equation}
For typical configurations with $R\ll 1$ and $|\mu|\ll |\lambda|$ we find that $dg_{\mathrm{eff}}/d\mu \approx -(g_{\mathrm{eff}}-\lambda_{\mathrm{eff}})^2 /(\lambda^2 R)$. 
Especially achieving large effective gains $g_{\mathrm{eff}} $ requires destructive quantum interference, i.e. $\lambda>0$ and $\mu<0$, and  the resulting small 
term $R\lambda +T\mu $ in the denominator of Eqs. (\ref{gformula}) and (\ref{geffformula}) implies high sensitivity of $g_{\mathrm{eff}}$ to changes of $\mu$.

Figure 4 illustrates the performance of teleportation-based noiseless quantum amplifier for two different gains. We consider noiseless amplification with pure entangled state (\ref{Psigeneralizedsubtraction}) 
and conditioning on $\beta=0$, which leads to the best possible performance of the amplifier. 
The results depicted in Fig.~4 thus provide a benchmark for comparison with predictions of more realistic models discussed in the next section. 
For small $\alpha$, the noiseless amplification is practically perfect. For larger $|\alpha|$, the resulting amplification gain and fidelity 
begins to decrease and the uncertainty product $V_x V_p$ grows,  indicating that the amplified state is no longer a minimum uncertainty state. 
Note that a deterministic amplifier would result in an uncertainty product $V_xV_p=g^2(\alpha)/2-1/4$, indicated by dashed lines in Fig. 4(d).

The conditional operation (\ref{Gdefinition}) approximates the noiseless amplifier well only for relatively weak coherent states with $|\alpha|<1$. 
Better approximation can be achieved with more sophisticated filters that approximate the unphysical operation $g^{\hat{n}}$ 
on a larger subspace of the Fock states. In particular, one can consider the following operation where the amplification is truncated at Fock state $|N\rangle$ \cite{Pandey2013,McMahon2014,Adnane2019},
\begin{equation}
\hat{G}_N=\frac{1}{g^N}\sum_{n=0}^N g^{n}|n\rangle\langle n|+\sum_{n=N+1}^\infty |n\rangle\langle n|.
\end{equation}
The corresponding normalized resource entangled state for teleportation-based noiseless amplification reads
\begin{equation}
|\Psi_N(\lambda)\rangle= \sqrt{\frac{1-\lambda^2}{P_N}}\left[\frac{1}{g^N}\sum_{n=0}^N (g\lambda)^{n} |n,n\rangle +\sum_{n=N+1}^\infty \lambda^n|n,n\rangle\right],
\label{PsiNdefinition}
\end{equation}
where
\begin{equation}
P_N=\frac{1}{g^{2N}}\frac{1-\lambda^2}{1-g^2\lambda^2}\left[1- (g\lambda)^{2N+2} \right]+\lambda^{2N+2}
\end{equation}
is a normalization factor. With the resource state (\ref{PsiNdefinition}) it is possible to achieve essentially perfect probabilistic noiseless amplification with finite probability $P_{\mathrm{tele}}>0$ 
for a range of coherent states with $|\alpha|<|\alpha_{\mathrm{th}}|$. The threshold amplitude $\alpha_{\mathrm{th}}$ can be estimated from the requirement that the effective support of the amplified 
coherent state $|g\lambda \alpha_{\mathrm{th}}\rangle$ falls within the subspace spanned by the first $N+1$ Fock states. 
We quantify the width of the coherent state in Fock basis by three standard deviations of its photon number distribution. 
This yields $g^2\lambda^2|\alpha_{\mathrm{th}}^2|+3g\lambda|\alpha_{\mathrm{th}}|=N$ and, consequently
\begin{equation}
|\alpha_{\mathrm{th}}|= \frac{1}{g\lambda}\left( \sqrt{N+\frac{9}{4}}-\frac{3}{2} \right).
\label{alphath}
\end{equation}

As a preparatory step for further discussion let us recall some facts about the continuous-variable teleportation of coherent states in the Braunstein-Kimble scheme \cite{Braunstein1998}. 
The measurement outcome $\beta$ indicates that the modes $A$ and $C$ were projected onto coherently displaced EPR state $\hat{I}_A\otimes \hat{D}_C(\beta^\ast)|\Psi_{\mathrm{EPR}}\rangle_{AC}$, where $\hat{D}(\beta)=\exp(\hat{a}^\dagger\beta-\hat{a}\beta^{\ast})$.
This can be equivalently interpreted as a coherent displacement of the input state in mode $C$,  $\hat{D}(\beta)|\alpha\rangle_C=|\alpha+\beta\rangle_C$,
followed by projection of modes $A$ and $C$ onto the EPR state $|\Psi_{\mathrm{EPR}}\rangle$. This in turn is equivalent to the projection of mode $A$ onto the re-normalized 
coherent state $\frac{1}{\sqrt{\pi}} |\alpha^\ast+\beta^\ast\rangle$.  The measurement outcomes $\beta$ span the whole complex plane and  the overcompleteness of projectors onto coherent states 
is accounted for by the prefactor $1/\sqrt{\pi}$ in our definition of the EPR  state $|\Psi_{\mathrm{EPR}}\rangle$, which ensures that we obtain correct probability density of measurement outcomes.

 Let us first consider teleportation with the Gaussian two-mode squeezed vacuum state (\ref{Psilambda}). Conditional on measurement outcome $\beta$, which occurs with probability density
\begin{equation}
P(\beta)=\frac{1}{\pi}(1-\lambda^2)e^{-(1-\lambda^2)|\alpha+\beta|^2},
\end{equation}
the mode $B$ is prepared in a coherent state $|\lambda\alpha+\lambda\beta\rangle$. If a corrective displacement operation $\hat{D}(-\lambda\beta)$ is applied to mode B, 
the teleported state becomes $|\lambda\alpha\rangle$, irrespective of the measurement outpcome $\beta$. 
The teleportation thus becomes equivalent to a purely lossy channel with amplitude transmittance $\lambda$ \cite{Fuwa2014}. 
This should be contrasted with the unity gain teleportation, where the corrective displacement operation reads $\hat{D}(-\beta)$ 
and the output state becomes a thermal dispaced state with coherent amplitude $\alpha$ and added thermal noise. 

Let us now turn to teleportation with the entangled state (\ref{PsiNdefinition}). If the coherent amplitude $\alpha+\beta$ satisfies the condition 
\begin{equation}
|\alpha+\beta| <|\alpha_{\mathrm{th}}|,
\label{betainequality}
\end{equation}
 then the  tail of the entangled state $|\Psi_N(\lambda)\rangle$ represented by the second sum in the formula (\ref{PsiNdefinition}) becomes irrelevant and the output state conditional on measurement outcome $\beta$ 
will be the noiselessly amplified coherent state $|g\lambda\alpha+g\lambda\beta\rangle$. After correcting displacement operation $\hat{D}(-g\lambda\beta)$ 
we recover the noislessly amplified input state $|g\lambda\alpha\rangle$. The probability density of the measurement outcomes $\beta$ can be expressed as 
\begin{equation}
P(\beta)= \frac{1}{\pi} \frac{1-\lambda^2}{g^{2N}P_N} e^{(g^2\lambda^2-1)|\alpha+\beta|^2},
\end{equation}
which is valid for $\beta$ that satisfy the inequality (\ref{betainequality}). In order to satisfy the condition (\ref{betainequality}) for some chosen range of input coherent amplitudes $\alpha$ 
the outcome of the teleportation-based amplifier is accepted only if $|\beta|$ is smaller than some threshold $\sigma$.  
In order to obtain analytical  expression for the resulting success probability $P_{\mathrm{tele}}$,
we can instead assume that the outcome $\beta$ is accepted with probability $\exp(-|\beta|^2/\sigma^2)$ and rejected otherwise. We have
\begin{equation}
P_{\mathrm{tele}}=\int_\beta P(\beta) e^{-\frac{|\beta|^2}{\sigma^2}} d^2\beta,
\label{pSintegral}
\end{equation}
which yileds
\begin{equation}
P_{\mathrm{tele}}=\frac{1-\lambda^2}{g^{2N}P_N} \frac{\sigma^2}{1+\sigma^2-\sigma^2\lambda^2g^2}\exp\left[\frac{(\lambda^2g^2-1)|\alpha^2|}{1+\sigma^2-\sigma^2\lambda^2g^2}\right].
\end{equation}
This formula is meaningful only if $\sigma$ and $|\alpha|$ are sufficiently small such that the following conditions are satisfied
\begin{equation}
\sigma^2 <\frac{1}{\lambda^2g^2-1},\qquad \frac{\lambda g|\alpha|}{1+\sigma^2-\sigma^2\lambda^2g^2}+\frac{3\lambda g \sigma}{\sqrt{2(\sigma^2+1-\sigma^2\lambda^2 g^2)}}  
< \sqrt{N+\frac{9}{4}}-\frac{3}{2} .
\label{sigmainequalities}
\end{equation}
The first inequality ensures that the Gaussian integral in (\ref{pSintegral}) converges and the second inequality guarantees that the range of complex amplitudes $\beta$ 
that non-negligibly contribute to the integral (\ref{pSintegral}) includes only $\beta$ for which the condition (\ref{betainequality}) is satisfied. 
The factor $3$ appearing in Eq. (\ref{sigmainequalities}) means that we take three standard deviations as the effective width of the involved Gaussian distribution of $\beta$. 

\section{Phase-space description of teleportation-based noiseless quantum amplifier}

In this section we present results of a more realistic model of the teleportation-based  noiseless quantum amplifier that takes into account the main experimental imperfections. 
The model is based on the well-established phase-space approach, where the Wigner function of a non-Gaussian state generated by photon subtractions from some Gaussian state can be expressed as a linear combination of
several Gaussian Wigner functions \cite{Fiurasek2005,Walschaers2020}. Here we choose to work with  the Husimi $Q$-function instead of the Wigner function, because with the $Q$-function 
it is particularly straightforward to calculate projections onto coherent states. 
Let us begin with some useful definitions. we consider system of $N$ modes, we denote by $\hat{x}_j$ and $\hat{p}_j$ the amplitude and phase quadrature operators of the $j$th mode, 
respectively, and and we collect the quadrature operators into a vector
 $\hat{\bm{q}}=(\hat{x}_1,\hat{p}_1,\ldots,  \hat{x}_N,\hat{p}_N)^T$. 
The covariance matrix $\gamma$ of the state is a $2N\times 2N$ square matrix  with elements
\begin{equation}
\gamma_{jk} =\langle \Delta\hat{q}_j\Delta\hat{q}_k+\Delta\hat{q}_k\Delta\hat{q}_j\rangle.
\end{equation}
where $\Delta \hat{q}_j=\hat{q}_j-\langle\hat{q}_j\rangle.$
In particular, a covariance matrix of a two-mode squeezed vacuum state (\ref{Psilambda}) with $\lambda=\tanh r$, where $r$ is the squeezing constant, reads
\begin{equation}
\gamma_{\mathrm{TMSV}}(r)=\left(
\begin{array}{cccc}
\cosh(2r) & 0 & \sinh(2r) & 0 \\
0 & \cosh(2r) & 0 & -\sinh(2r) \\
\sinh(2r) & 0 & \cosh(2r) & 0 \\
0 & -\sinh(2r) & 0 & \cosh(2r) 
\end{array}
\right).
\end{equation}
Note that the covariance matrix of vacuum state is equal to the identity matrix, $\gamma_{\mathrm{vac}}=I$.

Let us first describe the preparation of the two-mode resource state (\ref{Psigeneralizedsubtraction}) according to the scheme shown in Fig.~3. 
Our model assumes that the input states of modes $A,B$ and $C,D$ are Gaussian noisy two-mode squeezed thermal states that can be equivalently represented 
as initial pure two-mode squeezed states transmitted through a lossy channel with some effective transmittance $\eta$. The covariance matrices of the resulting mixed Gaussian states read
\begin{equation}
\gamma_{\mathrm{AB}}^{\mathrm{in}}=\eta_{\mathrm{AB}}\gamma_{\mathrm{TMSV}}(r)+(1-\eta_{\mathrm{AB}})I, 
\qquad \gamma_{\mathrm{CD}}^{\mathrm{in}}=\eta_{\mathrm{CD}}\gamma_{\mathrm{TMSV}}(s)+(1-\eta_{\mathrm{CD}})I,
\label{gammaTMSV}
\end{equation}
where $r$ and $s$ denote the bare squeezing constants of the initial pure states before losses, and  $\mu=\tanh s$. 
The parametrization in terms of effective losses is experimentally intuitive because it corresponds to a picture where the source generates pure states that are affected by losses both in the source and during the subsequent signal transmission. 

The single-photon detectors APD in Fig.~3 are modeled as on-off detectors  that can only distinguish the presence or absence of photons and have overall detection efficiency $\eta_{\mathrm{APD}}$. 
It is convenient to account for the inefficient detection  by letting the detected modes propagate through a lossy channel with transmittance $\eta_{\mathrm{APD}}$ followed by
 perfect on-off detectors whose click is described by POVM element $\hat{I}-|0\rangle\langle 0|$. In the state preparation scheme shown in Fig. 3 we condition on simultaneous clicks of both APDs 
 and the corresponding POVM element therefore reads
\begin{equation}
\hat{\Pi}_{CD}=(\hat{I}-|0\rangle\langle 0|)_C\otimes (\hat{I}-|0\rangle\langle 0|)_D.
\label{POVM}
\end{equation}
The effective covariance matrix of the overall Gaussian state of the four modes $A,B,C,D$ just before the measurement on modes $C$ and $D$ can be expressed as
\begin{equation}
\gamma_{\mathrm{eff,ABCD}}=S_{\mathrm{APD}} S_{\mathrm{BS}}(\gamma_{\mathrm{AB}}^{\mathrm{in}}\oplus \gamma_{\mathrm{CD}}^{\mathrm{in}})S_{\mathrm{BS}}^T S_{\mathrm{APD}}^T+G_{\mathrm{APD}},
\label{gammaeff}
\end{equation}
where
\begin{equation}
S_{\mathrm{PD}}= I_{\mathrm{AB}} \oplus \sqrt{\eta_{\mathrm{PD}}}I_{\mathrm{CD}}, \qquad G_{PD}= 0_{\mathrm{AB}}\oplus (1-\eta_{\mathrm{PD}})I_{\mathrm{CD}},
\end{equation}
represent the effective lossy channels in modes $C$ and $D$ that account for inefficient single-photon detection, and $S_{\mathrm{BS}}$
is a symplectic matrix that describes the coupling of modes $A,C$ and $B,D$ at the two beam splitters BS with transmittance $T$ and reflectance $R$. 

The Husimi $Q$-function of  a two-mode state $\hat{\rho}_{AB}$ is defined as
\begin{equation}
Q(\omega_A,\omega_B)=\frac{1}{\pi^2}\langle \omega_A,\omega_B|\hat{\rho}_{AB}|\omega_A,\omega_B\rangle,
\label{Qgeneraldefinition}
\end{equation}
where $|\omega_A,\omega_B\rangle$ denotes a coherent state of modes $A$ and $B$ with complex amplitudes $\omega_A$ and $\omega_B$, respectively. 
For Gaussian states $\hat{\rho}$ the $Q$-function (\ref{Qgeneraldefinition})  is a Gaussian distribution.
The POVM (\ref{POVM}) is a linear combination of four projectors onto Gaussian states. Therefore, the Husimi $Q$-function of the conditionally generated 
state of modes $A$ and $B$ can be expressed as linear combination of four Gaussian states \cite{Patron2004},
\begin{equation}
Q_{AB}(\omega_A,\omega_B)= \frac{1}{ P_{AB}}  \sum_{j=1}^{4} C_j  \sqrt{\frac{\det\tilde{\Gamma}_j}{\det\Gamma_j}} \frac{ \sqrt{\det \Gamma_j}}{\pi^2} e^{-\bm{r}_\omega^T\Gamma_j \bm{r}_\omega},
\label{QAB}
\end{equation}
where $C_1=C_4=1$, $C_2=C_3-1$, and $\bm{r}_\omega=[\Re(\omega_A),\Im(\omega_A),\Re(\omega_B),\Im(\omega_B)]^T$ is a real vector 
composed of real and imaginary parts of the complex amplitudes $\omega_A$ and $\omega_B$. 
The matrices $\tilde{\Gamma}_j$ have different sizes and are related to input covariance matrices of modes $AB$, $ABC$, $ABD$, and $ABCD$, respectively, 
\begin{equation}
\begin{array}{lcl}
\tilde{\Gamma}_{1}=2(\gamma_{\mathrm{eff}, AB}+I)^{-1}, &
\qquad  &
\tilde{\Gamma}_{2}=2(\gamma_{\mathrm{eff}, ABC}+I)^{-1}, \\
\tilde{\Gamma}_{3}=2(\gamma_{\mathrm{eff}, ABD}+I)^{-1}, &
\qquad &
\tilde{\Gamma}_{4}=2(\gamma_{\mathrm{eff}, ABCD}+I)^{-1}.
\end{array}
\end{equation}
Here the two-mode and three-mode covariance matrices $\gamma_{\mathrm{eff}, AB}$, $\gamma_{\mathrm{eff}, ABC}$ and $\gamma_{\mathrm{eff}, ABC}$ are submatrices of the full covariance matrix (\ref{gammaeff}). The matrices $\Gamma_j$ appearing in the exponent in Eq. (\ref{QAB})
are then extracted from  $\tilde{\Gamma}_{j}$ as two-mode submatrices corresponding to modes $A$ and $B$. Note that, in particular, $\Gamma_1=\tilde{\Gamma}_1$.
Finally, the probability $P_{AB}$ of conditional state preparation is given by
\begin{equation}
P_{AB}=\sum_{j=1}^4 C_j\sqrt{\frac{\det\tilde{\Gamma}_j}{\det\Gamma_j}}.
\end{equation}

With the phase-space representation of the resource entangled state we can proceed to analysis of the teleportation-based noiseless amplification of coherent states $|\alpha\rangle$. 
As discussed in the previous section,  the teleported state corresponding to measurement outcome $\beta$ can be obtained by projecting the mode $A$ of 
the two-mode entangled state (\ref{QAB}) onto the re-normalized coherent state  $\frac{1}{\sqrt{\pi}}|\alpha^\ast+\beta^\ast\rangle$. We introduce two real displacement vectors 
\begin{equation}
\bm{d}_\alpha=[\Re(\alpha),\Im(\alpha)]^T, \qquad \bm{d}_\beta=(\Re(\beta),\Im(\beta)]^T.
\end{equation}
To illustrate the calculations, consider teleportation with a generic two-mode Gaussian state with $Q$-function 
\begin{equation}
Q(\omega_A,\omega_B)=\frac{\sqrt{\det \Gamma}}{\pi^2} e^{-\bm{r}_{\omega}^T \Gamma \bm{r}_{\omega}}.
\label{QABGaussian}
\end{equation}
We introduce a bipartite $A|B$ splitting of the matrix $\Gamma$,
\begin{equation}
\Gamma=\left(
\begin{array}{cc}
\Gamma_A & M\\
M^T & \Gamma_B 
\end{array}
\right).
\end{equation}
The $Q$-function of a non-normalized state of  mode $B$ corresponding to projection of mode $A$ onto $\frac{1}{\sqrt{\pi}}|\alpha^\ast+\beta^\ast\rangle$ reads
\begin{equation}
Q_B(\omega_B|\beta)=\frac{\sqrt{\det\Gamma}}{\pi^2} \exp\left (-\bm{r}_{\omega_B}^T \Gamma_B \bm{r}_{\omega_B} 
-\bm{d}^T \Upsilon^T M \bm{r}_{\omega_B}-\bm{r}_{\omega_B}^TM^T \Upsilon \bm{d} -\bm{d}^T \Upsilon^T\Gamma_A \Upsilon \bm{d}\right) ,
\label{QBbeta}
\end{equation}
where $\bm{d}=\bm{d}_\alpha+\bm{d}_\beta$, $\bm{r}_{\omega_B}=[\Re(\omega_B),\Im(\omega_B)]^T$, and the matrix 
\begin{equation}
\Upsilon=\left(
\begin{array}{cc}
1 & 0 \\
0 & -1 
\end{array}
\right)
\end{equation}
accounts for the complex conjugation of $\alpha$ and $\beta$. Specifically, for $\beta=0$ we get
\begin{equation}
Q_B(\omega_B|\beta=0)= K(\alpha) \frac{\sqrt{\det\Gamma_B}}{\pi} \exp\left[-(\bm{r}_{\omega_B}-D\bm{d}_\alpha)^T\Gamma_B (\bm{r}_{\omega_B}-D \bm{d}_\alpha)\right],
\label{QBconditional}
\end{equation}
where
\begin{equation}
K(\alpha)=\frac{1}{\pi}\sqrt{\frac{\det\Gamma}{\det \Gamma_B}} e^{-\bm{d}_\alpha^T\tilde{\Gamma}_A\bm{d}_\alpha},
\end{equation}
and
\begin{equation}
\tilde{\Gamma}_A=\Upsilon^T\Gamma_A\Upsilon-\Upsilon^TM\Gamma_B^{-1}M^T\Upsilon, \qquad  D=-\Gamma_B^{-1}M^T\Upsilon.
\end{equation}
 The output Gaussian state (\ref{QBconditional}) has covariance matrix $\gamma_B=2\Gamma_B^{-1}-I$ and coherent displacement $D \bm{d}_\alpha$.

Let us now turn to teleportation with a non-Gaussian entangled resource  state (\ref{QAB}). We can write the normalized output teleported state conditioned on $\beta=0$ as a linear combination of four Gaussian states,
\begin{equation}
Q_{B}(\omega_B|\beta=0)= \frac{1}{ P_0}  \sum_{j=1}^{4} C_j \tilde{K}_j(\alpha)  \frac{\sqrt{\det\Gamma_{B,j}}}{\pi}e^{-(\bm{r}_{\omega_B}-D_j\bm{d}_\alpha)^T\Gamma_{B,j} (\bm{r}_{\omega_B}-D_j \bm{d}_\alpha)} ,
\label{QBbeta0}
\end{equation}
where
\begin{equation}
\tilde{K}_j(\alpha)=\frac{1}{\pi}\sqrt{\frac{\det\tilde{\Gamma}_j}{\det \Gamma_{B,j}}} e^{-\bm{d}_\alpha^T\tilde{\Gamma}_{A,j}\bm{d}_\alpha}, \qquad
P_0= \sum_{j=1}^{4} C_j \tilde{K}_j(\alpha).  
\end{equation}
The coherent displacement of the output state (\ref{QBbeta0}) can be calculated as
\begin{equation}
\bar{\bm{d}}=\frac{1}{P_0}\sum_{j=1}^{4} C_j \tilde{K}_j(\alpha)  D_j \bm{d}_\alpha. 
\end{equation}
The symmetry of the state-preparation scheme in Fig.~3 and the teleportation protocol in Fig.~2 ensures that for the considered class of the resource entanged states  (\ref{QAB}) 
the matrices $D_j$ are all proportional to the identity matrix, $D_j= d_j I$,  and the effective state-dependent amplification gain can be expressed as
\begin{equation}
g(\alpha)=\frac{1}{P_0}\sum_{j=1}^{4} C_j \tilde{K}_j(\alpha) d_j.
\label{gainformula}
\end{equation}
Following a similar procedure, one can also calculate the covariance matrix of the output  amplified state (\ref{QBbeta0}),
\begin{equation}
\gamma= \frac{1}{P_0}\sum_{j=1}^{4} C_j \tilde{K}_j(\alpha) \left (2\Gamma_{B,j}^{-1}+4D_j \bm{d}_\alpha \bm{d}_\alpha^T D_j^T\right) -4\bar{\bm{d}}\bar{\bm{d}}^T-I.
\label{cmformula}
\end{equation}
Finally, a fidelity of the output state (\ref{QBconditional}) with an amplified  coherent state $|g\alpha\rangle$ can be easily evaluated from the Husimi $Q$-function (\ref{QBconditional}),
\begin{equation}
\mathcal{F}(\alpha)= \pi Q_B(g\alpha|\beta=0)= \frac{1}{ P_0}  \sum_{j=1}^{4} C_j \tilde{K}_j(\alpha)  \sqrt{\det\Gamma_{B,j}} e^{-\bm{d}_\alpha^T(gI-D_j)^T\Gamma_{B,j} (gI-D_j) \bm{d}_\alpha} .
\label{fidelityformula}
\end{equation}

\begin{figure}[!t!]
\centerline{\includegraphics[width=\linewidth]{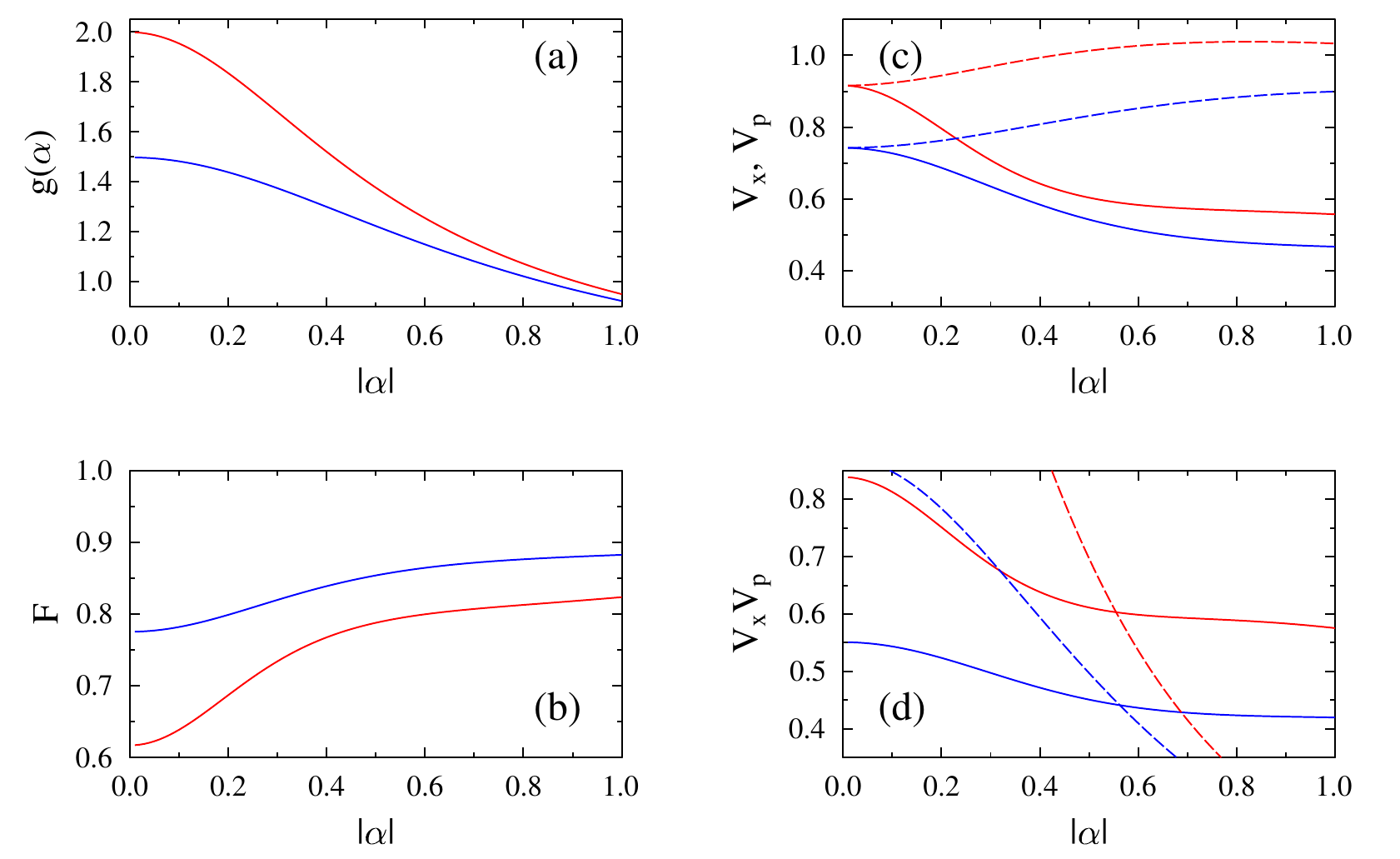}}
\caption{Performance of realistic noiseless amplifier with parameters  $\lambda=0.5$, $T=0.95$, $\eta_{\mathrm{APD}}=0.85$, $\eta_{AB}=\eta_{CD}=0.9$, and two different values of $\mu$ yielding effective gains $g_{\mathrm{eff}}=1.5$ ($\mu=-0.0150$, blue curves) 
and $g_{\mathrm{eff}}=2$ ($\mu=-0.0197$, red curves).  Conditioning on $\beta=0$ is assumed. The quantities plotted are the same as in Fig.~4: 
(a) amplitude-dependent amplification gain; (b) fidelity of the amplified state with coherent state $|g(\alpha)\alpha\rangle$, (c) variances 
of amplitude and phase quadratures of the amplified state; (d) product of quadrature variances of the amplified state.
}
\label{fig5}
\end{figure}

We have verified that if we consider pure input squeezed vacuum states ($\eta_{AB}=\eta_{CD}=1$), perfect detectors with $\eta_{\mathrm{APD}}=1$ and tapping beam splitters 
with very small reflectance $R<0.01$ , then for suitable choices of $\lambda$ and $\mu$ we recover the results predicted 
by pure state model and presented in the previous Section in Fig.~4. Let us now consider a more realistic scenario with noisy squeezed states, 
$\eta_{AB}=\eta_{CD}=0.9$ and superconducting single-photon detectors with total detection efficiency $\eta=0.85$.
We assume $\lambda=0.5$, which corresponds to $4$~dB of squeezing and $4.5$~dB of anti-squeezing in the resulting noisy 
two-mode squeezed thermal state with covariance matrix (\ref{gammaTMSV}). The results are plotted in Fig.~5 which displays the same quantities as Fig.~4. The results are again shown for two 
different values of the effective gain $g_{\mathrm{eff}}=1.5$ and $g_{\mathrm{eff}}=2$. The  values of squeezing strength $\mu$ that yield these gains were determined numerically and read $\mu=-0.0150$ and $\mu=-0.0197$, 
respectively. As expected, noisy squeezed states and imperfect single-photon detection with on-off detectors  
 lead to reduced performance of the noiseless amplifier in comparison to the ideal pure-state scenario considered in the previous section. Most notably, 
 the state fidelity is not a monotonously decreasing function of $|\alpha|$ but instead grows with $|\alpha|$. Note that this behavior is observed even if we consider 
fidelity with state $|g_{\mathrm{eff}}\alpha\rangle$ instead of $|g(\alpha)\alpha\rangle$. Similarly, the product of quadrature variances decreases with increasing $\alpha$.

To understand this behavior, we can consider setup with perfect photon number resolving detectors, vacuum in auxiliary modes $C$ and $D$ ($\mu=0$), 
and low thermal noise with mean number of thermal photons $\bar{n}_{\mathrm{th}}\ll 1$  in modes $A$ and $B$. In this limit, the amplified coherent state can be approximately 
expressed as a mixture of  three states $(\hat{n}+1)|\lambda\alpha\rangle$, $\alpha(\hat{n}+2)|\lambda\alpha\rangle$, and $(\hat{n}+1)\hat{a}^\dagger|\lambda\alpha\rangle$ 
where the probability of the last two states in the mixture is proportional to $\bar{n}_{\mathrm{th}}$. For $\alpha=0$ the state $\hat{a}^\dagger|\lambda\alpha\rangle$ 
becomes equal to the Fock state $|1\rangle$ that is orthogonal to the vacuum state. On the other hand, for $\alpha \neq 0$ the state $\hat{a}^\dagger|\lambda\alpha\rangle$ 
has a nonzero overlap with $|\lambda\alpha\rangle$ and this overlap increases with $|\alpha|$.  This explains the behavior of fidelity observed in Fig.~5(b).

In practice, a finite acceptance window for the outcomes $\beta$ will be required to achieve a non-zero overall success probability of the noiseless amplification. 
In order to obtain analytical results, we shall again assume that the outcomes $\beta$ are accepted with Gaussian probability $\exp(-|\beta|^2/\sigma^2)$. Simultaneously, 
we include a corrective coherent displacement  proportional to $\beta$, that is applied to the output state. For teleportation with entangled Gaussian state (\ref{QABGaussian}), 
the Husimi $Q$-function of the resulting output single-mode state can be expressed as
\begin{equation}
\tilde{Q}_B(\omega_B)=\int_\beta Q_B(\omega_B+k\beta|\beta)e^{-\bm{d}_\beta^T \Sigma \bm{d}_\beta} d^2\beta,
\label{QBintegral}
\end{equation}
where $\Sigma=\frac{1}{\sigma^2}I$  and $k$ is a constant that determines the strength of the corrective displacement. For Gaussian $Q$-function (\ref{QBbeta}) 
the integral in Eq. (\ref{QBintegral}) is a Gaussian integral that can be evaluated analytically. After a lengthy but straightforward calculation we obtain
\begin{equation}
\tilde{Q}_B(\omega_B)= K(\alpha) \frac{\sqrt{\det\tilde{\Gamma}_B}}{\pi} e^{-(\bm{r}_{\omega_B}-\tilde{D}\bm{d}_\alpha)^T\tilde{\Gamma}_B (\bm{r}_{\omega_B}-\tilde{D} \bm{d}_\alpha)},
\label{QBtildesigma}
\end{equation}
where
\begin{equation}
K(\alpha)=\frac{1}{\sqrt{\det\Gamma_\beta}}\sqrt{\frac{\det\Gamma}{\det \tilde{\Gamma}_B}} e^{-\bm{d}_\alpha^T\tilde{\Gamma}_A\bm{d}_A}.
\label{Ksigma}
\end{equation}
In order to write down the formulas for the various matrices appearing in Eqs. (\ref{QBtildesigma}) and (\ref{Ksigma}) we first specify the matrix $\Gamma_\beta$,
\begin{equation}
\Gamma_\beta=k^2\Gamma_B+\Upsilon^T\Gamma_A\Upsilon+\Sigma+k\Upsilon^TM+kM^T\Upsilon,
\end{equation}
and define two auxiliary matrices
\begin{equation}
L_\alpha=\Upsilon^T\Gamma_A \Upsilon +k M^T \Upsilon,  \qquad L_r=\Upsilon^T M+k \Gamma_B.
\end{equation}
With the help of these definitions we can write down explicit formulas for the remaining matrices,
\begin{eqnarray}
& \tilde{\Gamma}_B= \Gamma_B-L_r\Gamma_\beta^{-1}L_r, & \nonumber \\[2mm]
&\tilde{\Gamma}_A=\Upsilon^T\Gamma_A\Upsilon-L_\alpha^T\Gamma_\beta^{-1}L_\alpha-   (\Upsilon^TM -L_\alpha^T\Gamma_\beta^{-1}L_r) \tilde{\Gamma}_B^{-1}(M^T\Upsilon-L_r^T\Gamma_\beta^{-1}L_\alpha), 
& \\[2mm]
&\tilde{D}=-\tilde{\Gamma}_B^{-1}(M^T\Upsilon-L_r^T\Gamma_\beta^{-1}L_\alpha).& \nonumber
\end{eqnarray}

If we apply the above Gaussian integration to each of the four terms in the formula (\ref{QAB}) for the $Q$-function of the photon subtracted two-mode squeezed state, 
we obtain the final expression for the $Q$-function of the teleported state,
\begin{equation}
Q_{B}(\omega_B)= \frac{1}{ P_{\mathrm{tot}}}  \sum_{j=1}^{4} C_j \tilde{K}_j(\alpha)  \frac{\sqrt{\det\tilde{\Gamma}_{B,j}}}{\pi}
e^{-(\bm{r}_{\omega_B}-\tilde{D}_j\bm{d}_\alpha)^T\tilde{\Gamma}_{B,j} (\bm{r}_{\omega_B}-\tilde{D}_j \bm{d}_\alpha)} . 
\label{QBbetasigma}
\end{equation}
Here 
\begin{equation}
\tilde{K}_j(\alpha)=\frac{1}{\sqrt{\det\Gamma_{\beta,j}}}\sqrt{\frac{\det\tilde{\Gamma}_j}{\det \tilde{\Gamma}_{B,j}}} e^{-\bm{d}_\alpha^T\tilde{\Gamma}_{A,j}\bm{d}_A}, \qquad P_{\mathrm{tot}}
= \sum_{j=1}^{4} C_j \tilde{K}_j(\alpha).
\label{Ktildesigma}
\end{equation}
and $P_{\mathrm{tot}}$ is the total success probability that is a product of the probability of state preparation $P_{AB}$ and probability of successful conditional teleportation 
$P_{\mathrm{tele}}$. Therefore, $P_{\mathrm{tele}}=P_{\mathrm{tot}}/P_{AB}$. The same machinery that was described above for conditioning on $\beta=0$ can be used to calculate 
the various properties of the teleamplified state (\ref{QBbetasigma}) such as the effective gain, state fidelity or quadrature variances, by employing the analogues of Eqs. (\ref{gainformula}), 
(\ref{cmformula}) and (\ref{fidelityformula}).

\begin{figure}[!t!]
\centerline{\includegraphics[width=\linewidth]{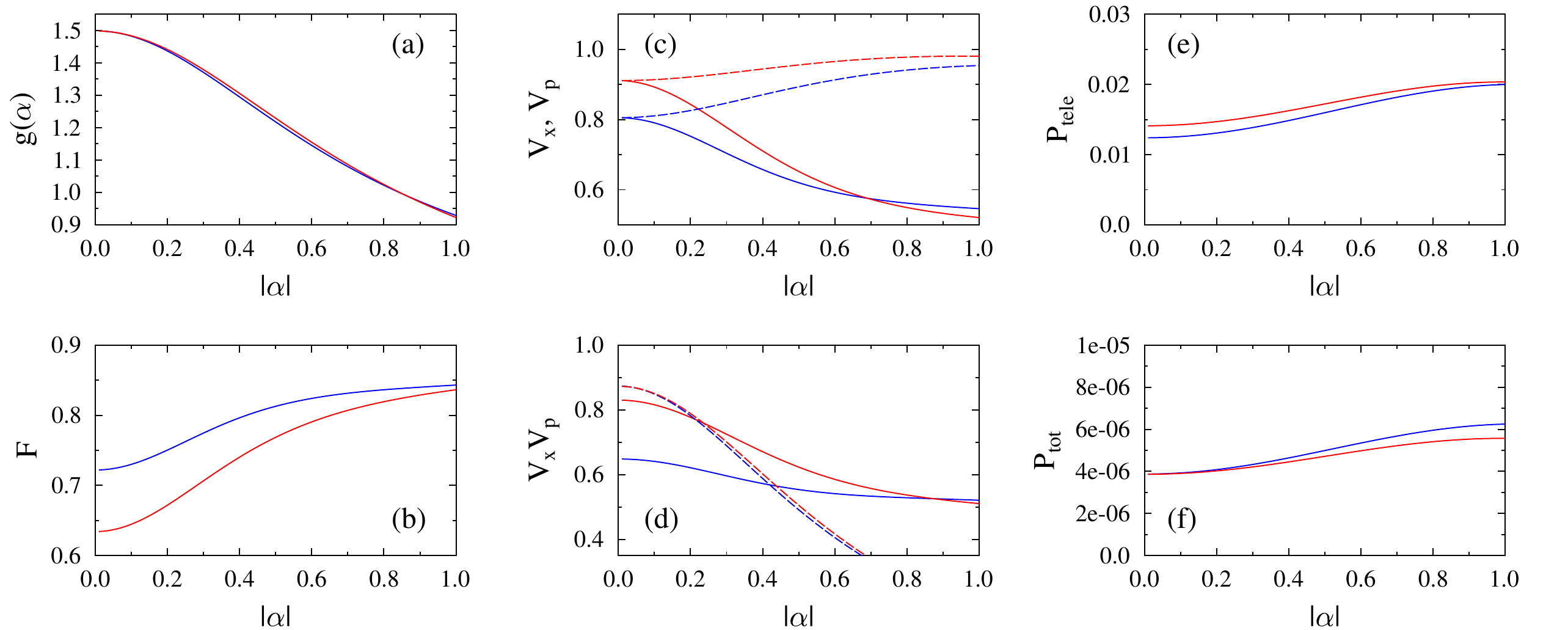}}
\caption{Performance of realistic noiseless amplifier with parameters  $\lambda=0.5$,  $\eta_{\mathrm{APD}}=0.85$, $\eta_{AB}=\eta_{CD}=0.9$, $g_{\mathrm{eff}}=1.5$,  
and finite acceptance window for measurement outcomes $\beta$, $\sigma^2=0.08$.  The blue curves indicate results for a protocol with corrective displacement: $k=1$, $T=0.95$, $\mu=-0.0179$. 
For reference, the red curves display results for a protocol without any corrective displacement: $k=0$, $T=0.955$, and $\mu=-0.01475$. The panels (a)-(d) display the same quantities as in Figs.~4 and 5. 
In the two additional panels (e,f) we plot the success probabilities $P_{\mathrm{tele}}$ (e), and $P_{\mathrm{tot}}$ (f).}
\label{fig3}
\end{figure}

Figure 6 illustrates the performance of the teleportation-based noiseless amplifier for a finite acceptance window of measurement outcomes $\beta$ ($\sigma^2=0.08$). 
We compare the protocols with ($k=1$) and without ($k=0$) corrective coherent displacement. 
To make a fair comparison, we choose the auxiliary squeezing parameter $\mu$ and the transmittance $T$ of the tapping beam splitters 
separately for each case  to achieve the same effective gain $g_{\mathrm{eff}}=1.5$ and the same total success probability $P_{\mathrm{tot}}$ for small $\alpha$. 
We can see that the gain curves practically overlap, but the protocol with corrective displacement achieves higher fidelity and lower quadrature fluctuations of the amplified state.  
These results therefore clearly demonstrate the usefulness and advantage of the corrective displacement. As shown in Fig. 6(e), the success probability of  the teleportation protocol 
itself is of the order of $1\%$ for the parameters considered. The total success probability plotted in Fig. 6(f) is mainly reduced by small probability of preparation of the photon subtracted two-mode entangled state, here
$P_{AB} \approx 3\times 10^{-4}$. As discussed above, the state preparation can be accomplished before the teleamplification is attempted 
and in this sense the probability $P_{\mathrm{tele}}$ characterizes the probabilistic noiseless teleamplifier. However, with current technology it would not be possible to prepare the entangled state 
in advance and store it in a quantum memory while maintaining the required high quality of the state. Practical implementation of the protocol would therefore 
require simultaneous success of the conditional state preparation and teleportation, which is characterized by the total probability o success $P_{\mathrm{tot}}$.

\section{Conclusions}
In summary, we have proposed and theoretically analyzed a scheme for  noiseless quantum amplification of coherent states of light via probabilistic quantum teleportation 
with a suitably designed two-mode entangled state. With this approach we can implement noiseless amplifiers similar to those based on combination of conditional single photon addition 
and subtraction while avoiding the experimentally costly photon addition operation. Instead, the protocol requires auxiliary two-mode squeezed vacuum states and balanced homodyne detection 
in addition to the conditional single photon subtraction. In its simplest version, the noiseless amplification is achieved with a two-mode squeezed vacuum with single photon subtracted from 
each of its modes. However, the achievable gain $2\lambda$ is limited by the available squeezing and cannot exceed $2$. More control over the amplification gain can be obtained 
by utilizing an additional second two-mode squeezed vacuum state with low squeezing $\mu$. With this extended scheme, arbitrary high gain (for low $|\alpha|$) is in principle achievable, 
but the setup may become sensitive to fluctuations of  $\mu$. Therefore, the simplest variant with $\mu=0$ is experimentally most feasible. 

The teleportation-based noiseless amplifier requires high purity squeezing to operate properly. During recent years we have witnessed tremendous progress in generation of highly pure 
strongly squeezed states of light, see. e.g. \cite{Suzuki2006,Mehmet2011,Vahlbruch2016,Asavanant2017,Mehmet2019}. For instance, in Ref. \cite{Vahlbruch2016}, 
generation of squeezed state with $10$~dB of squeezing and only $11$~dB of anti-squeezing is reported. 
In our numerical simulations, we have considered a considerably lower squeezing of about $4$~dB for which even higher purity is achievable. 
Imperfections of single photon detectors mainly limit the success probability of the protocol, while the quality of the noiseless amplifier is preserved 
if beam splitters with sufficiently low reflectance $R$ are employed. With recent significant development of superconducting single photon detectors, 
quantum detection efficiencies exceeding $90\%$ become available, with recently reported system detection efficiency of $98\%$ \cite{Reddy2020}. 
Furthermore, the spatial or temporal multiplexing can be used to turn on-off detectors into approximate photon number resolving detectors \cite{MeyerScott2020}, 
which in combination with higher reflectances $R$ of the tapping beam splitters 
could significantly increase the success probability of photon subtraction while preserving high quality of the condiitonally generated state. 

Finally, we would like to point ou that the teleportation of quantum gates can be applied also to other important elementary conditional continuous-variable 
quantum operations such as single photon addition. This requires as a resource state a two-mode squeezed vacuum state with a single photon subtracted from mode A,
\begin{equation}
\hat{a}|\Psi_{\mathrm{TMVS}}(\lambda)\rangle=\lambda\sqrt{1-\lambda^2} \sum_{n=0}^\infty \sqrt{n+1}\lambda^n |n,n+1\rangle_{AB}.
\end{equation}
For arbitrary input state $|\psi\rangle_C$, conditioning on $\beta=0$ in the teleportation scheme in Fig.~3, the conditionally prepared output state in mode $B$ 
becomes $\hat{b}^\dagger \lambda^{\hat{n}}|\psi\rangle_B$. Similarly to the teleportation-based noiseless amplification, conditioning on finite range of 
measurement outcomes $\beta$ will unavoidably introduce some noise and imperfections into the implemented conditional operation. Nevertheless, 
the effective replacement of photon addition by photon subtraction can make this approach appealing for certain applications.

\section*{Funding}
Czech Science Foundation (21-23120S).

\section*{Disclosure}
The authors declare that there are no conflicts of interest related to this article.

\end{document}